# 3.55 keV Silicon Anomaly as X-ray Candle in the Universe

V.V. Burdyuzha

**Nuclear physics and astrophysics department of P.N. Lebedev Physical Institute, Russian Academy of Sciences, Moscow 117997, Russia**

The Ly-α line of $^{14}Si$ with the energy 3.55-keV is formed in the magnetic field $6·10^{12}$ G. In close binary stellar systems with a red giant and a neutron star in super-strong magnetic fields recombination laser radiation to the Landau ground level may appear from hydrogen-like ions. If the 3.55-keV Ly-α silicon laser realizes, then the detection of this line from any distances in the Universe on $z \leq 100$ is possible and this line may be the X-ray candle because of a huge abundance of silicon in these binary systems. Energies for a Ly-α line of hydrogen-like $^{14}Si$ in magnetic fields $4·10^{12}$-$10^{13}$ G are calculated. This line is the unique event because we see this radiation from perpendicular direction to the magnetic column in near-wall layers. The laser radiation to the ground level in the range of (1-20) keV may arise from other hydrogen-like ions. In the presence of a laser, line narrowing is inevitable.

**Kew words: recombination, laser, super strong magnetic fields, neutron stars, red giants**

## 1. Introduction

In Space the thermo-dynamical equilibrium is absent and a non-equilibrium radiation may arise practically everywhere. The short introduction to Space masers was published by Burdyuzha (2014). All observed masers radiate in radio or millimeter range. Of course the non-equilibrium radiation in Space must take place in other diapasons also. In this article we consider possibility of realization of laser radiation in the X-ray range based on electron transition in hydrogen-like atoms in an extremely strong magnetic field. Recently some emission X-ray lines were observed by (Boyarsky et al, 2014; 2019; Bulbul et al, 2014; Iakubovskyi et al, 2015; Cappelluti et al, 2018; Gall et al, 2019 ) in keV range. The 3.55-keV line was observed in the spectrum of the Andromeda Galaxy, in the spectrum of some clusters of galaxies with red shifts z=0.01-0.35, and in the direction to the center of our Galaxy (Jeltema & Profumo, 2015). It was called the 3.55-keV anomaly because of the absence of its reliable identification. This line was discovered by several X-ray satellites: XMM-Newton, Chandra, Suzaku and NuSTAR. 3.55-keV anomaly called a large interest due to its interpretation through photons from decaying dark matter on sterile neutrino (Boyarsky et al, 2014, 2019; Cline et al, 2014; Lovell et al, 2019). But some skepticism arises to its explanation by decaying dark matter: Jeltema & Profumo (2016); Riemer-Sorensen (2016); Perez et al, (2017); Dessert, Rodd & Safdi (2018); Ng et al, (2019). Besides, X-ray lines with energies higher and lower than 3.55-keV were also observed (Boyarsky et al, 2014, Bulbul et al, 2014).

Here we show that the 3.55-keV line and lines near can appear in close binary systems consisting of a red giant and a neutron star in the super strong magnetic fields. The laser radiation in the 3.55-keV line takes place. Probably, this is an observational fact because sensitivity of the X-ray detector was not sufficient to detect the 3.55-keV line in thermal regime on the red shift z ~0.35. The 3.55-keV Ly-α line can be related to recombination on hydrogen-like $^{14}$Si taking place in the near-wall layers of the magnetic column of a neutron star upon efficient cooling (Burdyuzha 2019 (a)). Of course, recombination with emission of X-ray Ly-α quantum can also occur on other hydrogen-like ions: $^2$He, $^6$C, $^7$N, $^8$O, $^{10}$Ne, $^{12}$Mg, $^{16}$S, $^{20}$Ca, $^{26}$Fe.

**2. Hydrogen-like ions in super-strong magnetic fields of neutron stars**

Because of the cylindrical symmetry in super strong magnetic fields, the energies of excitations coinciding with the field direction and perpendicular to it are different and three quantum numbers denote the energy levels. Then, transitions to the Landau ground level are denoted as $E_{001} \rightarrow E_{000}$ (along the field) and $E_{0-10} \rightarrow E_{000}$ (perpendicular to the field) where $E_{000}$ is the Landau ground level. Quantum numbers of any energetic level ($E_{vmn}$) in super-strong magnetic fields denote: ν – number of Landau level; m-projection of moment, n- parity of a level (Burdyuzha and Pavlov-Verevkin, 1981). Here, we present a few simple expressions used for calculations and give transition energies. Recall that magnetic fields in which the radius of the first Bohr orbit exceeds the cyclotron radius of an electron are called super-strong magnetic fields (SMF):

$$a_Z = \hbar^2 / e^2 Z^{3/2} > \rho_o = (c\hbar/eB)^{1/2} \qquad (1)$$

For hydrogen, the super strong magnetic field B≥B$_0$ is realized already for

$$B_0 = m_e^2 c e^3 / \hbar^3 = 2.35 \cdot 10^9 \, G \qquad (2)$$

For hydrogen-like ions, this condition has the form (Kadomtsev1970):

$$B > B_0 Z^2 \qquad (3)$$

All the details about boundary conditions, formulation of the problem, and the finite-difference method used for exact calculations of energy levels in SMF are presented by Burdyuzha & Pavlov-Verevkin (1981). Here the range of magnetic fields $2 \cdot 10^{11} - 10^{13}$ G covers the majority of X-ray pulsars. In the magnetic field range $4 \cdot 10^{12}$-$10^{13}$ G our calculations for hydrogen-like $^{14}$Si gave the following energies (in eV) for the $E_{0-10} \rightarrow E_{000}$, $E_{001} \rightarrow E_{000}$ transitions which are collected in the Table 1.

| B (G) | $4 \cdot 10^{12}$ | | $6 \cdot 10^{12}$ | | $8 \cdot 10^{12}$ | | $10^{13}$ | |
|---|---|---|---|---|---|---|---|---|
| | $E_{0-10} \rightarrow E_{000}$ | $E_{001} \rightarrow E_{000}$ | $E_{0-10} \rightarrow E_{000}$ | $E_{001} \rightarrow E_{000}$ | $E_{0-10} \rightarrow E_{000}$ | $E_{001} \rightarrow E_{000}$ | $E_{0-10} \rightarrow E_{000}$ | $E_{001} \rightarrow E_{000}$ |
| $^{14}$Si | 3180 | 6860 | **3550** | 8090 | 3840 | 9070 | 4080 | 9920 |

**Table 1.** Energy levels in eV of H-like silicone in magnetic fields of $4 \cdot 10^{12}$-$10^{13}$ G

For calculation of the energetic levels we used the non-relativistic Schrodinger equation describing the motion of an electron in a magnetic field and in the field of an infinitely heavy nucleus with charge Z. In atomic units ($\hbar = e = m_e = 1$) it is:

$$\{-(1/2)(\partial^2/\partial\rho^2) - (1/2)(\partial^2/\partial z^2) + (1/8)[(4m^2-1)\rho^{-2} + B^2\rho^2] - (Z/\sqrt{\rho^2 + z^2})\} \psi(z, \rho) =$$

$$= (E - (1/2) B\, m) \psi(z, \rho) \qquad (4)$$

Here m -projection of the angular momentum of an electron, z-direction along of magnetic field, ρ-direction perpendicularly to magnetic field. Note, that in magnetic fields only the transverse motion of the charged particles is quantized. Also we must estimate the luminosity of the Ly-α line for hydrogen-like silicon in the magnetic column of a neutron star in the thermal regime. Probabilities for the Ly-α line of hydrogen-like ions in a huge magnetic field are proportional to $Z^4$:

$$A^i = A^H Z^4 \, s^{-1} \qquad (5)$$

For hydrogen, $A^H \approx 10^9$ s$^{-1}$ and for Si, $A^{Si} \approx 4 \cdot 10^{13}$ s$^{-1}$ for Z=14. Then, the maximum luminosity of the Ly-α line of Si in the magnetic field $6 \cdot 10^{12}$ G may be:

$$L^{Si} = EnVA = 3.55 \cdot 10^3 \times 1.6 \cdot 10^{-12} \times 10^{19} \times 5 \cdot 10^{15} \times 4 \cdot 10^{13} \approx 10^{40} \text{ erg s}^{-1} \qquad (6)$$

Here E-energy of transition; n-density; V-volume; A-probability of transition. The diameter of the magnetic column in this estimate was $10^5$ cm, the internal diameter was $6 \cdot 10^4$ cm, the column height was $10^6$ cm and then the volume was V~$5 \cdot 10^{15}$ cm$^3$. But in a magnetic column of a neutron star in close binary systems the abundance of silicon ions differs essentially because silicon ions produce in atmospheres of red giants abundantly. They are the cores of interstellar dusts after them blowing from atmospheres of these stars. For calculations we took n=$10^{19}$ cm$^{-3}$ although the inequality ($H^2/8\pi$) > nkT admits to have the density up to n=$10^{29}$ cm$^{-}$ 3 at T~$10^{10}$ K. We cannot determine exactly the density of silicone ions and the estimate in (6) may be overestimated even. The energies and luminosities of other hydrogen-like ions in magnetic fields from $10^{12}$ to $10^{13}$ G can be obtained from expressions (5) and (6) for the transition $E_{0-10} \to E_{000}$.

| $^2$He | $^6$C | $^7$N | $^8$O | $^{10}$Ne | $^{12}$Mg | $^{14}$Si | $^{16}$S | $^{20}$Ca | $^{26}$Fe |
|---|---|---|---|---|---|---|---|---|---|
| 199 | 1030 | 1290 | 1570 | 2180 | 2840 | **3550** | 4310 | 5990 | 8840 |
| <$10^{36}$ | $10^{39}$ | $5 \cdot 10^{38}$ | $6 \cdot 10^{39}$ | $3 \cdot 10^{39}$ | $4 \cdot 10^{39}$ | $10^{40}$ | $10^{40}$ | $5 \cdot 10^{39}$ | $3 \cdot 10^{41}$ |

**Table 2.** Energy levels and luminosities in the field of $6 \cdot 10^{12}$ G for $E_{0-10} \to E_{000}$

Here for calculations the space abundance of elements was used. The full table of energies for these ions in magnetic fields of $(4-10) \cdot 10^{12}$ G was presented by Burdyuzha (2019(a)). Abundance of $^2$He ions is difficult to define.

## 3. Laser radiation of 3.55 keV line from the magnetic column of neutron stars

A separate question is appearance of laser radiation in that magnetic column of a neutron star in the near-wall layers. An intensifying medium of any plasma is characterized by its constant aggregate state for a high density of input energy. This medium allows an effective amplification radiation in the short wavelengths. Many years ago Gudzenko and Shelepin (1963) drew attention to the recombining plasma as an active medium. The plasma starts recombining due to cooling which may be accompanied by emission (radiation cooling). During cooling the recombination flux populates the upper "working" levels of ions in the plasma and an inverse population of these levels with respect to the lower ones occurs. This concept was realized in the 80's in the Livermore and Princeton laboratories (Rosen et al., 1985, Suckewer et al., 1985). The laser effect in the X-ray range was obtained in plasma by a nuclear explosion (Ritson, 1987) also. The possibility of recombination lasing at the 3.55-keV line and the lines of other ions was also predicted by Burdyuzha & Pavlov-Verevkin (1981). In the two-level $N_b \to N_a$ model, amplification can occur at frequency $\omega_{ba}$. The population inversion condition has the form:

$$\delta_{ab} \equiv (N_a/g_a) / (N_b/g_b) < 1 \qquad (7)$$

The gain in the unsaturated recombination regime is described by a simple expression:

$$\varkappa = \sigma_{ab} N_e (1 - \delta_{ab}) \qquad (8)$$

Here: $\sigma_{ab} = (\lambda_{ab}^2/4) A_{ab}/\Delta\omega_{ab}$ is the absorption cross section at the line center, $N_e$ is the electron density, $\lambda_{ab}$ is the amplified radiation wavelength, and $\Delta\omega_{ab}$ is the line width. In the magnetic column of a neutron star, we have a specific case when X-ray lasing can appear only in near-wall layers where the Landau ground level is efficiently cooled. One can easily see that the Landau ground level $E_{000}$ is depleted due to the two $E_{001} \leftarrow E_{000}$ and $E_{0-10} \leftarrow E_{000}$ radiation transitions and collisions. For $n_e \sim 10^{19}$ cm$^{-3}$ $\sigma_{ab} \sim 5\times 10^{-20}$ cm$^2$ is available. In the unsaturated recombination lasing regime, the high value of $\varkappa l$ can be obtained even for small population inversion ~0.05%, the low electron density ~$10^{19}$ cm$^{-3}$, and the wall thickness of the magnetic column $l \sim 4\times 10^4$ cm. The exponential factor can achieve the value $\varkappa l=10$, and the Ly-$\alpha$ line will be amplified $e^{10}$ times, i.e. 22 000 times. If exponential factor $\varkappa l=12$ then the 3.55-keV line may be amplified in 170 000 times. These estimates are rough and only illustrative because the balance equation was not solved. But the important additional moment is necessary to note. In the unsaturated regime of laser radiation a narrowing of this X-ray line in $\sqrt{\varkappa l}$ times should be that is 3.1-3.5 times. Of course, the same is right for X-ray lasing in other hydrogen-like ions: $^2$He, $^6$C, $^7$N, $^8$O, $^{10}$Ne, $^{12}$Mg, $^{16}$S, $^{20}$Ca, and $^{26}$Fe. More probable emission of Ly-$\alpha$ lines near the line 3550 eV in close binary systems may appear from ions of $^7$N (1290 eV), $^8$O (1570 eV), $^{10}$Ne (2180 eV), $^{12}$Mg (2840 eV), $^{16}$S (4310 eV), $^{20}$Ca (5990 eV), $^{26}$Fe (8840 eV). They must have widths which are not similar to the width of the 3.55 keV line. In bright X-ray pulsars, radiation pressure comes into play and decelerates the accreting material. Because of that, there are huge gradients of velocity in the accretion column. And these gradients of velocity should result in a line broadening (Lipunov, 1987; Shakura, 2016). Any lines have to be affected by the Doppler effects, of course. Ly-$\alpha$ line of $^{10}$Ne, $^{12}$Mg, and $^{16}$S were probably observed by Boyarsky et al. (2014).

## 4. Observation capabilities

The geometry of a magnetic column of neutron stars is complex (we do not always see the perpendicular component or see it weakened at an angle). The longitudinal component is affected by the gravitational red shift (the ordinary Doppler shift takes place for both components). The emission component exactly along the magnetic field is not observed practically due to the channel geometry. Besides, the red shift is not so small ($\Delta E \sim 0.1$ keV for $M_{NS}=2M_\odot$). More detail hydrodynamics and geometry of accretion columns of neutron stars in binary stellar systems was considered by Basko and Sunyaev (1976) a long time ago.

For comparison the energies for Ly-α transition of H-like ions in absence of a magnetic field are: $^6$C →367.5 eV, $^7$N →500.3 eV, $^8$O →653.6 eV; $^{10}$Ne →1022 eV, $^{12}$Mg →1472 eV, $^{14}$Si →2006 eV, $^{16}$S →2622 eV, $^{20}$Ca →4105 eV, $^{26}$Fe→ 6976 eV. Of course, there is a strong anisotropy for all other X-ray lines also. Naturally, other lines are less intense due to the huge silicon abundance but a forest of weak lines must be observed. In article of Riemer-Sorensen (2016) some lines in the range (2-9) keV were identified from Chandra observations of Galactic Center. These were lines of different ions. In addition, the region in keV contains an electron cyclotron line:

$$\hbar\omega_{cycl} \approx 1.2(B/10^{11}) \{1- [GM/R]/c^2]\} \quad \text{keV} \quad (9)$$

Occurrence of the electron cyclotron line is more likely than rise of Ly-α line from any elements. If B=6·10$^{12}$ G, then ℏω$_{cycl}$ = 72 keV. A fresh review of cyclotron line research was recently published by Staubert, Trumper, Kendziorra, et al. (2019).

## 5. Conclusion

Many years ago some important notes of magnetized plasma in Space were done by Gnedin, Pavlov, and Tsygan (1974). Owing to the high value of the photo-ionization cross-section in magnetic fields 10$^{12}$-10$^{13}$ G, photo-recombination can be the principal source of release of thermal energy by plasma of neutron stars at temperature T>10$^6$-10$^7$ K. The observed 3.55-keV Ly-α line (silicon anomaly) could be emitted during recombination of hydrogen-like silicon in a close binary system consisting of a neutron star and a red giant in the super strong magnetic field 6·10$^{12}$ G. Such magnetic field may be realized in the magnetic column of a neutron star. Recombination occurs to the Landau ground state of $^{14}$Si and other ions in near-wall layers. Highlighting the 3.55 keV line is connected with the huge abundance of silicon ions that enter the magnetic column. Modern X-ray telescopes already at present can discover the 3.55-keV recombination line from the most remote "silicon sources" in the Universe for any red shifts z ≤ 100. This line is the unique event (Burdyuzha, 2019(a)) because we see this radiation from perpendicular direction to the magnetic column. At high red shifts it will be the UV diapason already. Besides, 3.55-keV Ly-α line may be the Universe X-ray candle. Ly-α lines from other hydrogen-like ions $^2$He, $^6$C, $^7$N, $^8$O, $^{10}$Ne, $^{12}$Mg, $^{16}$S, $^{20}$Ca, $^{26}$Fe in SMF of close binary stellar systems must be detected in the range (1-20) keV (Burdyuzha & Pavlov-Verevkin, 1981; Burdyuzha, 2019(a)). Due to these new possibilities, the X-ray spectroscopy should receive the additional impulse to the new launching Japanese Hitomi satellite and new X-ray satellites. Note that in the abstract of our full article (Burdyuzha, 2019(a)) the typographic error took place: instead of $10^{12}$G was $10^{22}$G. This error was corrected (Burdyuzha, 2019 (b)).

The author thanks A. V. Tutukov from Institute of Astronomy of Russian Academy of Sciences in Moscow for consultations on close binary stars.